\RequirePackage{fix-cm}
\documentclass[twocolumn]{svjour3}          
\smartqed  
\usepackage{graphicx}
\usepackage[caption=false,font=normalsize,labelfont=sf,textfont=sf]{subfig}
\usepackage{lineno,hyperref}
\usepackage{amsmath,graphicx}
\usepackage{amssymb}
\usepackage{algpseudocode,algorithm}
\usepackage{graphicx}
\usepackage{epstopdf}
\usepackage{balance}
\usepackage{graphicx}
\usepackage{epstopdf}
\usepackage{times,cite}
\usepackage{textcomp}
\graphicspath{{./figuresFinalToSumbit/}}
\usepackage[caption=false,font=normalsize,labelfont=sf,textfont=sf]{subfig}
\graphicspath{{./figures/}}
\begin{document}

\title{V-Shaped Sparse Arrays For 2-D DOA  Estimation 
}


\author{Ahmet M. Elbir 
}


\institute{ Ahmet M. Elbir \at
              Dept. of Electrical and Electronics Engineering, Duzce University, 81620, Duzce, Turkey
              \email{ahmetmelbir@gmail.com, ahmetelbir@duzce.edu.tr}           
}

\date{Received: date / Accepted: date}

\maketitle

\begin{abstract}
This paper proposes a new sparse array geometry for 2-D (azimuth and elevation) DOA (direction-of-arrival) estimation. The proposed array geometry is V-shaped sparse array and it is composed of two linear portions which are crossing each other. The degrees of freedom of the sparse array is enhanced by sparse sampling property. In this respect, V-shaped coprime (VCA) and V-shaped nested array (VNA) structures are developed. VCA can resolve both azimuth and elevation angles up to $ MN$ sources with $2M + N -1$ sensors in each portion  and the total number of sensors is $4M+2N-3$. VNA can resolve $O(N^2)$ sources with $2N$ sensors. Instead of 2-D grid search, the proposed method computes 1-D search for azimuth and elevation angle estimation in a computational efficient way. In order to solve the pairing problem in 2-D scenario, the cross-covariance matrix of two portion is utilized and 2-D paired DOA estimation is performed. The performance of the proposed method is evaluated with numerical simulations and it is shown that the proposed array geometries VCA and VNA can provide much less sensors as compared to the conventional coprime planar arrays.
\end{abstract}
	\section{Introduction}
{D}{irection-of-arrival} (DOA) estimation is an important issue in array signal processing for a number of applications such as radar, sonar and wireless communications \cite{kitap}. The MUSIC (MUltiple SIgnal Classification) algorithm \cite{music} is one of the most powerful methods in this context due to its simplicity and asymptotic performance with respect to the corresponding performance bounds. The effectiveness of the MUSIC algorithm is attributed to the orthogonality of signal and noise spaces and the performance limit of the MUSIC algorithm is to estimate up to  $K\leq M-1$ source directions for an $M$-element sensor array since at least one-dimensional noise subspace is required.

While in most of the applications uniform array structures are used \cite{friedlander}, nonuniform arrays \cite{nonUniformArray1,nestedArray,nonUniformArray2,nonUniformArray3,coprimeDSPConf} gain much interest recently due to their efficiency in terms of number of sensor elements and providing underdetermined source estimation where there are more sources than sensors, i.e. $M < K$. In earlier studies, nonuniform array structures are considered in the context of array interpolation in \cite{ref_AI5,ref_AI6,ref_AI7}, however the property of the array to handle underdetermined scenario is not exploited. One of the nonuniform array structures is the minimum redundancy arrays (MRAs) which are discussed in \cite{nonUniformArray1}. While MRA provides higher degrees of freedom (DOF) than usual uniform linear arrays (ULAs), there is no closed for expression to obtain the sensor positions of an MRA for a certain number of sensors $M$ \cite{nestedArray}. In \cite{nonUniformArray2,nonUniformArray3}, the augmentation of covariance matrices for enhancing DOF is proposed where the resulting covariance matrix is not positive semidefinite for a finite number of snapshots. In \cite{nestedArray}, nested array structures are proposed for estimating $O(M^2)$ sources with $O(M)$ sensors. Since nested arrays have more closely spaced sensors which eventually cause relatively higher mutual coupling, coprime array structures are introduced in \cite{coprimeDSPConf} where the array is composed of less number of element pairs that are closely spaced and hence less coupling occurs. Using a 1-D coprime array, up to $K \leq MN$ sources can be identified with only $2M + N -1 $ sensor elements. Note that above array structures are 1-dimensional (1-D) and they cannot be employed for 2-D (azimuth and elevation) DOA estimation. 

2-D DOA estimation using coprime arrays are considered in \cite{Coprime2DPlanarSJ,LshapedCoprimeGeneralizedCL2017}. 
The authors in \cite{Coprime2DPlanarSJ} propose a coprime planar array (CPA) structure. In particular, CPA consists of $M_1\times M_1$ and $M_2\times M_2$ subarrays where $M_1$ and $M_2$ are coprime integers. It is reported that this method can resolve $K \leq \text{min} \{M_1^2,M_2^2\} -1$ sources with $M_{\text{CPA}}= M_1^2 + M_2^2$ sensor elements. The method in \cite{LshapedCoprimeGeneralizedCL2017} generalizes the construction of coprime planar arrays (GCPA) and \cite{LshapedCoprimeGeneralizedCL2017} uses $N_1\times M_1$ and $N_2 \times M_2$ two subarrays where $N_1,N_2$ and $M_1,M_2$ are coprime integer sets. Hence GCPA can resolve $K \leq \text{min}\{N_1M_1,N_2M_2\}-1$ sources which provides higher DOF than CPA using $M_{\text{GCPA}} = N_1M_1 + N_2M_2$ sensors.

For 2-D DOA estimation, instead of planar arrays, L-shaped arrays provide much simpler structure and it is widely used for 2-D DOA estimation \cite{LshapedSparseRsFinding1,LshapedNestedPairedIET,LshapedCrossCorrAWPL2016,LshapedApertureSnapshotExtentionAWPL2016,LshapedApertureSnapshotExtentionSPL2017}. In \cite{LshapedSparseRsFinding1}, steering matrix estimation is done for the estimation of azimuth and elevation angles separately. In \cite{LshapedNestedPairedIET}, L-shaped nested arrays are considered for the same problem. In \cite{LshapedCrossCorrAWPL2016, LshapedApertureSnapshotExtentionAWPL2016, LshapedApertureSnapshotExtentionSPL2017}, augmented data matrices are constructed for aperture and snapshot extension to utilize the structure of L-shaped arrays. While L-shaped array is a promising choice for 2-D DOA estimation\cite{LshapedSubmittedSPL}, it returns coupled estimation results \cite{nielsenUncoupledDOA,tansuVshaped}. In other words, the azimuth and elevation angles are coupled and the error in estimation of one parameter (say azimuth) affects the accuracy of the other (elevation). Hence a more general array geometry is required for uncoupled 2-D DOA estimation.

\begin{figure*}[t]
	\centering
	\subfloat[]{\includegraphics[width=.25\textheight,height=.17\textheight]{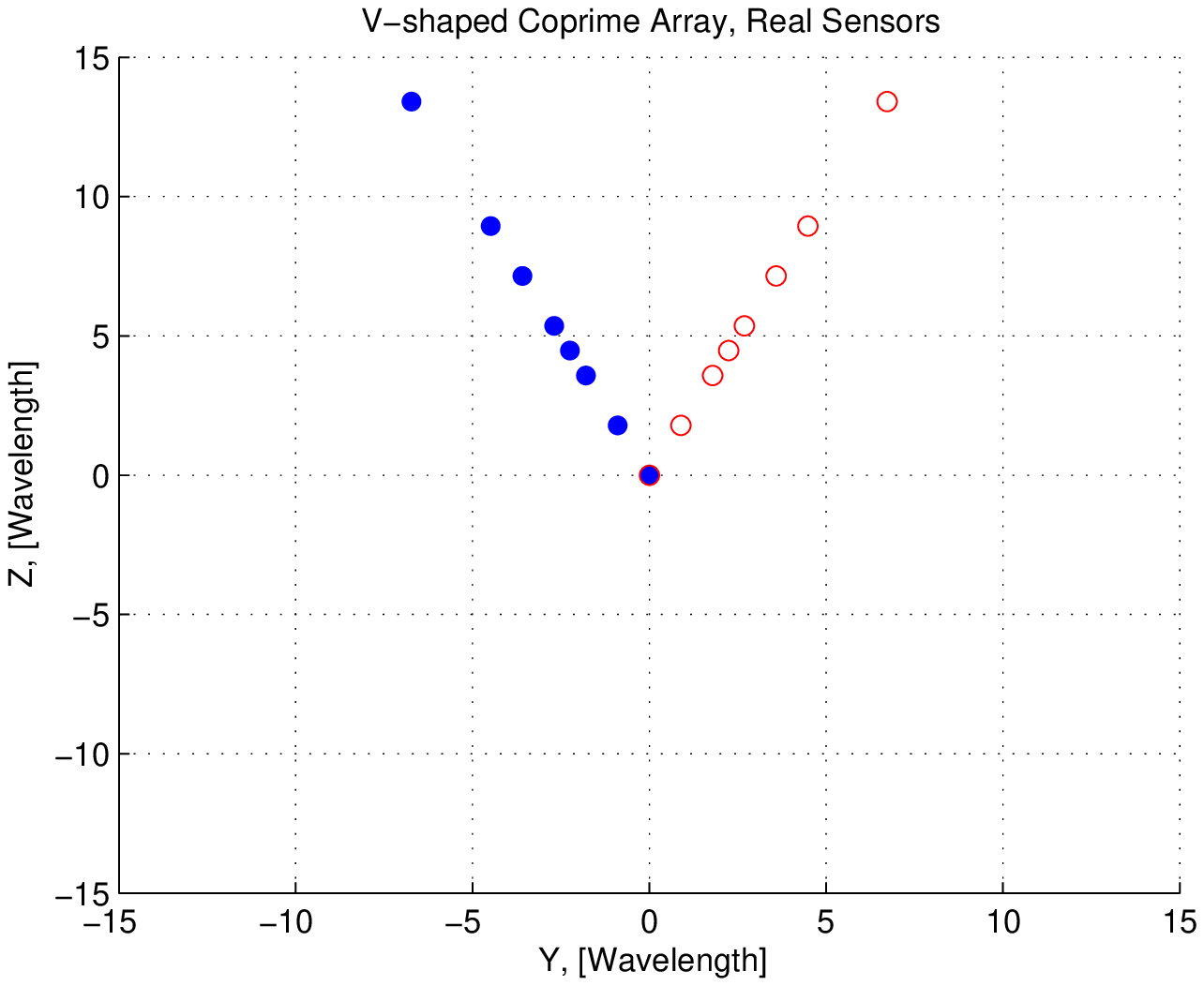}%
		\label{figPosa} } 
	\subfloat[]{\includegraphics[width=.25\textheight,height=.17\textheight]{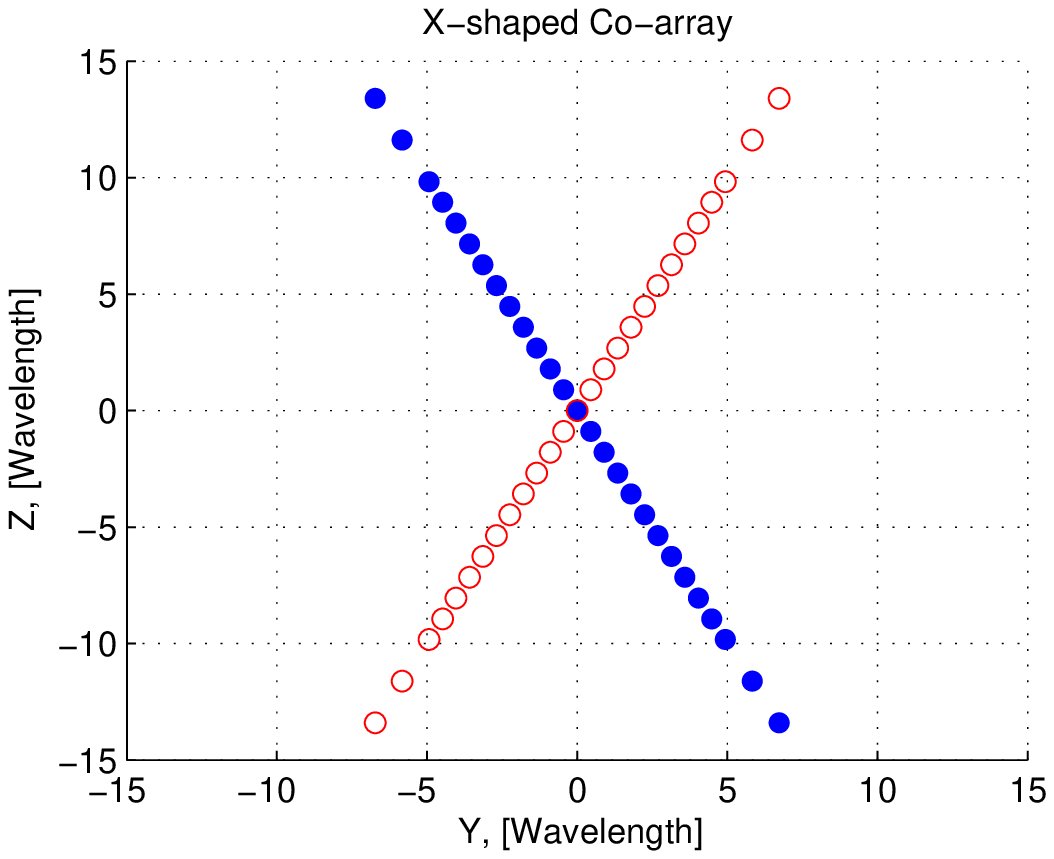}%
		\label{figPosb}	}
	\subfloat[]{\includegraphics[width=.25\textheight,height=.17\textheight]{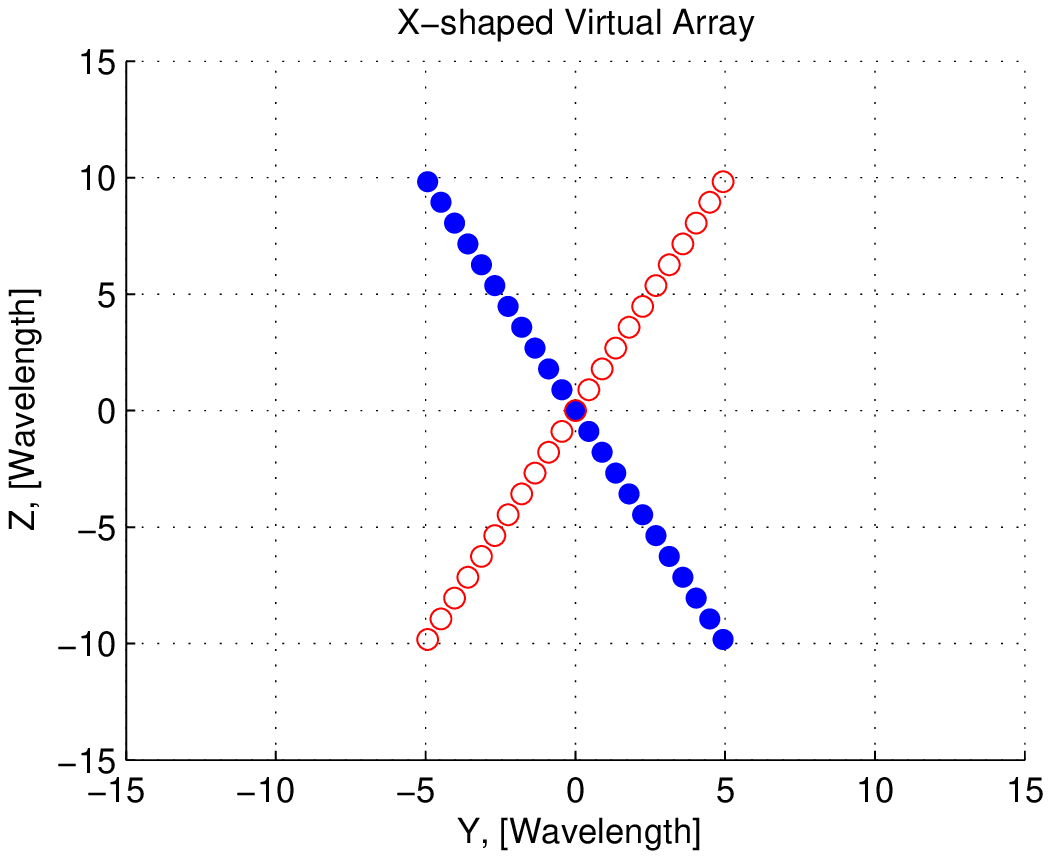}%
		\label{figPosc}	}
	\caption{V-shaped coprime array (VCA) structure for $M = 2$, $N = 5$ and $d = \lambda/2$. (a) The real sensor positions. (b) Co-array of each portion of VCA. (c) The contagious part of each co-array.}
	\label{figPos}
\end{figure*}
Instead of L-shaped arrays, V-shaped sparse arrays (VSAs) provide more flexibility and generalizes the concept of L-shaped arrays. Moreover, V-shaped arrays can be designed so that uncoupled DOA estimation is achieved. In this paper, a new sparse array geometry, V-shaped coprime array (VCA), is proposed for 2-D DOA estimation. The proposed array geometry is composed of two portions in two axis, namely, $\mathcal{U}$- and $\mathcal{V}$-axes (See Sec. \ref{secArrayModel} for their definitions). In each portion, there are $2M+N-1$ sensors to incorporate the coprime sparse arrays. Hence the total number of sensors in the array is $M_{\text{VCA}}=4M+2N-3$. VCA provides $O(MN)$ DOF in each portion. This leads to the fact that VCA can resolve $K\leq MN$ sources. The proposed V-shaped array geometry is also extended to nested arrays and V-shaped nested arrays (VNAs) are constructed using only $2N$ sensors and it provides $O(N^2)$ DOF. We show that the proposed array structures, VCA and VNA, provide much less sensor elements as compared to other 2-D nonuniform arrays such as CPA \cite{Coprime2DPlanarSJ} and GCPA \cite{LshapedCoprimeGeneralizedCL2017}. In the proposed DOA estimation technique, firstly the design of VSA is considered and the V-angle of VSA which leads to uncoupled DOA estimation is obtained. In order to estimate the 2-D DOA angles the sparse structure of each portion is utilized and a longer virtual ULA is constructed by vectorization of the covariance matrix of data from each portion. Since the obtained data model is in Vandermonde form, spatial smoothing is employed then the rank-enhanced covariance matrix is obtained \cite{coprimeDSPConf,spatialSmoothingRemark}. The covariance matrices of each portions in $\mathcal{U}$- and $\mathcal{V}$-axes  are used for azimuth and elevation angle estimation respectively. In order to obtain paired 2-D DOA angles, the cross-covariance matrix between the data of each portion is used and automatically paired 2-D DOA estimation is achieved. The major contributions of the proposed method are as follows:
\begin{enumerate}
	\item The proposed array structures, VCA and VNA, require much less sensor elements as compared to the other planar sparse arrays, CPA and GCPA, for 2-D parameter estimation.
	\item VSA geometries provide uncoupled 2-D DOA estimation which enables to obtain accurate results when estimating azimuth and elevation angles separately.
	\item The proposed method does not require 2-D search method which is computational inefficient and it can simply be performed using 1-D search algorithms such as the MUSIC algorithm to obtain the line spectra for azimuth and elevation separately.
\end{enumerate}

The remainder of the paper is as follows. In Sec. \ref{secArrayModel}, the array model is presented for coprime case and relevant details are provided. Sec. \ref{secDesignV} studies the design of the VSA and the computation of the V-angle. In Sec. \ref{secDOACoprime}, DOA estimation for coprime arrays is introduced for azimuth and elevation angles separately. In Sec. \ref{secPaired2D}, the proposed paired 2-D DOA estimation algorithm is introduced. In Sec. \ref{secNested}, the proposed approach is extended for nested arrays. Sec. \ref{secCompComp} considers the computational complexity of the proposed method. The numerical simulations are presented in Sec. \ref{Simulations} and in Sec. \ref{secConc}, the paper is finalized with conclusions. 
\begin{figure}[t]
	\centering
	{\includegraphics[width=.30\textheight]{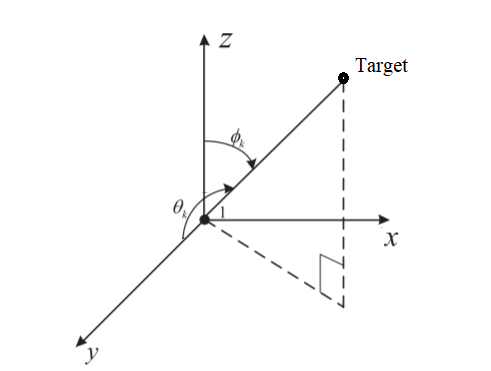} } 
	\caption{The definition of elevation and azimuth angles, $(\theta,\phi)$.}
	\label{coordinateSystem}
\end{figure}
\section{Array Signal Model}
\label{secArrayModel}
Consider a V-shaped array composed of two portions placed in $yz$-plane as seen in Fig \ref{figPos}a. For simplicity, the axes on which the sensors are placed are called $\mathcal{U}$- and $\mathcal{V}$-axes. $\mathcal{U}$-axis is defined for the sensors with the position set $\mathbb{U} = \{y_i,z_i: -y_i\sin(\Omega/2) + z_i\cos(\Omega/2) = u_i \}$ where $y_i,z_i$ are the sensor positions in Cartesian coordinate system and $\Omega$ is the V-angle between two portions. $\mathcal{V}$-axis is also defined in a similar way for $\mathbb{V} = \{y_i,z_i: y_i\sin(\Omega/2) + z_i\cos(\Omega/2) = v_i\}$.  Note that $u_i$ and $v_i$ are integer numbers which contribute to enhance the aperture of the sparse array and obtain larger virtual array with coprime property and Vandermonde model. Each portion consists of two subarrays with $2M$-  and $N$-elements where $M < N$ and $M,N \in \mathbb{N}^{+}$ are coprime numbers \cite{coprimeDSPConf}. The locations of the $2M$ sensors are in the set $\mathbb{S}_{2M} = \{Nmd: 0\leq m \leq 2M-1 \}$ and the locations of the $N$ sensors are in the set $\mathbb{S}_{N}=\{Mnd : 0\leq n \leq N-1 \}$ respectively where $d$ is the fundamental element spacing in the array and $d=\lambda/2$ for narrowband source signals to avoid spatial aliasing \cite{stoicaBook}. Therefore there are $2M+N-1$ sensors in each portion and the total number of sensors in the array is $M_{\text{VCA}} = 4M + 2N -3$. Assume that there are $K$ source signals impinging on the array from directions $\Theta_k = \{\theta_k,\phi_k\}_{k=1}^K$  where $\theta_k$ and $\phi_k$ are being the elevation and the azimuth angle of the $k$th source respectively (See Fig.~\ref{coordinateSystem}). Then the outputs of each portion are given by
\begin{align}
\label{sigModely}
\ \boldsymbol{\mathcal{U}}(t_i) =& \sum_{k=1}^{K} \textbf{a}_{u}(\Theta_k) {{s}}_k(t_i) + \ {\textbf{n}_{{\mathcal{U}}}}(t_i), \\ 
\label{sigModelz}
\boldsymbol{\mathcal{V}}(t_i) =& \sum_{k=1}^{K} \textbf{a}_{v}(\Theta_k) {{s}}_k(t_i) + \ {\textbf{n}_{{\mathcal{V}}}}(t_i)
\end{align}
where $i = 1,\dots,T$  and $T$ is the number of snapshots and $\ \textbf{n}_{u}(t_i),\textbf{n}_{v}(t_i) \in\mathbb{C}^{(2M+N-1)}$ are temporarily and spatially white noise vectors.  $\{{s}_k(t_i)\}_{k=1,i=1}^{K,T}$ is the set of uncorrelated source signals and $\textbf{a}_{u}(\Theta_k),\textbf{a}_{v}(\Theta_k)$ denote the steering vectors corresponding to the $k$th source and their $i$th elements are given by
\begin{align}
\label{aUandaV}
\left[{\textbf{a}_{u}}(\Theta_k)\right] _i= & {\text{exp}} \{j\frac{2\pi }{\lambda} {u_i}[-\sin(\phi_k) + \sin(\theta_k)]  \} \\
\label{aUandaV2}
\left[{\textbf{a}_{{v}}}(\Theta_k)\right] _i =& {\text{exp}} \{j\frac{2\pi }{\lambda}  {v_i}[\sin(\phi_k) + \sin(\theta_k)]\}
\end{align}
where $[\cdot]_i$ denotes $i$th element of the vector quantity. ${u_i}$ and ${v_i}$ are the sensor positions in $\mathcal{U}$- and $\mathcal{V}$-axes respectively. $\lambda$ is the wavelength and $u_i,v_i \in \mathbb{S}$ which is defined as $\mathbb{S} =  \mathbb{S}_{2M} \cup \mathbb{S}_{N}$, i.e.
\begin{eqnarray}
\mathbb{S} = 	\left\lbrace Mnd : 0\leq n \leq N-1 \right\rbrace \cup \left\lbrace Nmd :0\leq m \leq 2M-1   \right\rbrace . \nonumber
\end{eqnarray}

Note that the azimuth and elevation angles in (\ref{aUandaV}) and (\ref{aUandaV2}) are  defined different than the conventional definition as in \cite{LshapedApertureSnapshotExtentionAWPL2016} and \cite{LshapedNestedPairedIET} (i.e., $\theta $ and $\phi$ are the angles between the source and the $y$- and $z$-planes respectively in Fig.~\ref{coordinateSystem}). Hence a unique transformation between each other can always be performed without loss of generality.

The aim in this work is to estimate DOAs $\{\theta_k,\phi_k\}_{1\leq k \leq K} $ of $K \leq MN$ ($K\leq N^2/4+N/2-1$ for VNA) sources by using only $M_{\text{VCA}} = 4M + 2N-3$ ($2N$ for VNA) sensors when the sensor positions $\{u_i,v_i \}_{1\leq i \leq 2M+N-1}$ and $\Omega$ are known.

\textit{Remark:} Due to the computation of the noise subspace in the MUSIC algorithm, the proposed method requires the knowledge of the number of sources $K$. While the estimation process of $K$ is an exclusive work and an important issue in many fields of array signal processing \cite{numberOfSourceEst,numberOfSourceEst2}, in this letter it is assumed that $K$ is known \textit{a priori}.

\section{Design Of V-Shaped Coprime Array}
\label{secDesignV}
The design of V-shaped array includes the determination of the sensor positions in accordance with coprime sampling property and the determination of the V-angle $\Omega$. While it seems $\Omega$ has no major effect in DOA estimation, it determines the coupling between the azimuth and elevation angle estimation. For a certain value of $\Omega$, the azimuth and elevation angle estimation problems are uncoupled, i.e. the estimation error of azimuth (or elevation) does not affect the accuracy of elevation (azimuth) estimation. In order to obtain uncoupled DOA estimation, the cross terms of the Fisher information matrix need to be zero \cite{tansuVshaped,nielsenUncoupledDOA}. This condition can be satisfied by placing the sensors in accordance with the V-angle selected as  $\Omega = 2\text{tan}^{-1}\{\sqrt{\frac{\bar{M}^2 + 3}{4\bar{M}^2}} \}$ where $\bar{M} = 2MN+1$ is the number of sensors in virtual V-shaped array \cite{tansuVshaped}. For $M=2,N = 5$, $\bar{M} = 21$ and $\Omega = 53.28^{\circ}$ as shown in Fig. \ref{figPos}.

\section{DOA Estimation With Coprime Arrays}
\label{secDOACoprime}
Using the array model in (\ref{sigModely}) and (\ref{sigModelz}), the covariance matrices for each portion are defined as
\begin{align}
\label{Ry}
{\textbf{R}}_{\boldsymbol{\mathcal{U}}} =& E \{ \boldsymbol{\mathcal{U}}(t)  \ \boldsymbol{\mathcal{U}}^H(t)\} = \textbf{A}_{u} \textbf{R}_{\text{S}} \textbf{A}_{u}^H + \sigma_n^2 \textbf{I}, \\
\label{Rz}
{\textbf{R}}_{\boldsymbol{\mathcal{V}}} =& E \{\boldsymbol{\mathcal{V}}(t) \ \boldsymbol{\mathcal{V}}^H(t)\} = \textbf{A}_{v} \textbf{R}_{\text{S}} \textbf{A}_{v}^H + \sigma_n^2 \textbf{I},
\end{align}
where $\textbf{A}_{u}$ and $\textbf{A}_{v}$ are $(2M+N-1) \times K$ steering matrices whose $k$th columns are $\textbf{a}_{u}(\Theta_k)$ and $\textbf{a}_{v}(\Theta_k)$ respectively. $\textbf{R}_{\text{S}} = \text{diag}\{\sigma_1^2,\dots,\sigma_K^2\}$ is  $K \times K$ signal correlation matrix, $\textbf{I}$ is the identity matrix and $\sigma_n^2$ is the noise variance.

Due to the structure of coprime arrays, a longer virtual array can be constructed by taking advantage of the second order statistics  $\textbf{R}_{\boldsymbol{\mathcal{U}}}$ and $\textbf{R}_{\boldsymbol{\mathcal{V}}}$. While the real array includes the lags given in the set $\mathbb{S}$, the elements of $\textbf{R}_{\boldsymbol{\mathcal{U}}}$ and $\textbf{R}_{\boldsymbol{\mathcal{V}}}$ can provide a larger position set whose elements constitute the difference co-array $\mathbb{S}_{\text{diff}}$ which is defined as the unique terms in the set
\begin{eqnarray}
\mathbb{S}^2 = \left\lbrace (Mn -Nm)d : 0\leq n \leq N-1, 0\leq m \leq 2M-1   \right\rbrace \nonumber.
\end{eqnarray}
In Fig. \ref{figPos}b, the difference co-array of each portion of the VCA is presented. In order to exploit the co-array structure inherit in the covariance matrices, vectorization is applied to $\textbf{R}_{\boldsymbol{\mathcal{U}}}$ and $\textbf{R}_{\boldsymbol{\mathcal{V}}}$ and we get
\begin{align}
\boldsymbol{\mathcal{U}}_{\mathbb{S}^2} =& \text{vec} \{ \textbf{{R}}_{\boldsymbol{\mathcal{U}}} \} 
= {\textbf{{A}}}_{{u}_{{\mathbb{S}^2}}} \textbf{{p}} + {\textbf{{n}}}_{u_{{\mathbb{S}^2}}}, \\
\boldsymbol{\mathcal{V}}_{\mathbb{S}^2} =& \text{vec} \{ \textbf{{R}}_{\boldsymbol{\mathcal{V}}} \}
= {\textbf{{A}}}_{{v}_{{\mathbb{S}^2}}} \textbf{{p}} + {\textbf{{n}}}_{{v}_{{\mathbb{S}^2}}},
\end{align}
where ${\textbf{{A}}}_{{u}_{{\mathbb{S}^2}}} = {\textbf{{A}}}_{u}^* \circledcirc {\textbf{{A}}}_{u}$, ${\textbf{{A}}}_{{v}_{{\mathbb{S}^2}}} = {\textbf{{A}}}_{v}^* \circledcirc {\textbf{{A}}}_{v}$ and $\circledcirc$ denotes the Khatri-Rao product\cite{khatriRaoProduct,coprimeDSPConf}.	$\textbf{{p}} = [\sigma_1^2,\dots, \sigma_K^2]^T$  represents the signal powers and ${\textbf{{n}}}_{u_{{\mathbb{S}^2}}} = {\textbf{n}}_{v_{{\mathbb{S}^2}}} = \text{vec}\{ \sigma_n^2 \textbf{{I}}\}$. Now observe that $\boldsymbol{\mathcal{U}}_{{{\mathbb{S}^2}}}$ can be viewed as the output of a virtual array with  sensor positions $\mathbb{S}_{\text{diff}}$ which includes $2MN+1$ contiguous terms from $-MN$ to $MN$ as seen in Fig. \ref{figPosc}. Hence a longer virtual ULA can be constructed from the row elements of ${\boldsymbol{\mathcal{U}}}_{{{\mathbb{S}^2}}}$, say ${\boldsymbol{\mathcal{U}}}_{\mathbb{S}_{\text{diff}}^{\text{ULA}}}$, i.e.
\begin{eqnarray}
{\mathbb{S}_{\text{diff}}^{\text{ULA}}} = \left\lbrace nd: -MN\leq n \leq MN  ,  \right\rbrace 
\end{eqnarray}
is the set of sensor positions of $(2MN+1)$-element virtual ULA. Therefore the rows of ${\boldsymbol{\mathcal{U}}}_{\mathbb{S}^2}$ and ${\boldsymbol{\mathcal{V}}}_{\mathbb{S}^2}$ corresponding to ${\mathbb{S}_{\text{diff}}^{\text{ULA}}}$ are collected and the following virtual array data for a single snapshot is obtained, i.e.
\begin{eqnarray}
\label{virtualULA1}
\boldsymbol{\mathcal{U}}_{\mathbb{S}_{\text{diff}}^{\text{ULA}}} = {\textbf{A}}_{{u}_{\mathbb{S}_{\text{diff}}^{\text{ULA}}}} \textbf{p} + {\textbf{n}}_{{u}_{\mathbb{S}_{\text{diff}}^{\text{ULA}}}}, \\
\boldsymbol{\mathcal{V}}_{\mathbb{S}_{\text{diff}}^{\text{ULA}}} = {\textbf{A}}_{{v}_{\mathbb{S}_{\text{diff}}^{\text{ULA}}}} \textbf{p} + {\textbf{n}}_{{v}_{\mathbb{S}_{\text{diff}}^{\text{ULA}}}},
\end{eqnarray}
where $ {\textbf{A}}_{{u}_{\mathbb{S}_{\text{diff}}^{\text{ULA}}}}, {\textbf{A}}_{{v}_{\mathbb{S}_{\text{diff}}^{\text{ULA}}}} \in \mathbb{C}^{(2MN+1) \times K}$ are the array manifold matrices corresponding to the sensor elements with positions $u_i,v_i \in {\mathbb{S}_{\text{diff}}^{\text{ULA}}}$. In order to estimate the DOA angles, the MUSIC algorithm can be applied to the covariance matrices of ${\boldsymbol{\mathcal{U}}}_{\mathbb{S}_{\text{diff}}^{\text{ULA}}}$ and ${\boldsymbol{\mathcal{V}}}_{\mathbb{S}_{\text{diff}}^{\text{ULA}}}$. Since the resultant covariances will be rank 1, spatial smoothing is required to estimate the DOA angles. In order to obtain a spatially smoothed covariance matrix a rank-enhanced Toeplitz positive semidefinite matrix is constructed  \cite{spatialSmoothingRemark} where the smoothed covariance matrix is obtained from the observations ${\boldsymbol{\mathcal{U}}}(t_i)$ and ${\boldsymbol{\mathcal{V}}}(t_i)$ directly.  Then the smoothed covariance matrix $\textbf{R}_{\boldsymbol{\mathcal{U}}-\text{SS}}$ is obtained as
\begin{eqnarray}
\textbf{R}_{\boldsymbol{\mathcal{U}}-\text{SS}} = \left( \begin{array}{llll}
\left[	\tilde{\boldsymbol{\mathcal{U}}}_{\mathbb{S}_{\text{diff}}^{\text{ULA}}}\right]_L & \left[\tilde{\boldsymbol{\mathcal{U}}}_{\mathbb{S}_{\text{diff}}^{\text{ULA}}}\right]_{L-1} & \dots & \left[\tilde{\boldsymbol{\mathcal{U}}}_{\mathbb{S}_{\text{diff}}^{\text{ULA}}}\right]_{1} \\  
\left[\tilde{\boldsymbol{\mathcal{U}}}_{\mathbb{S}_{\text{diff}}^{\text{ULA}}}\right]_{L+1} & \left[ \tilde{\boldsymbol{\mathcal{U}}}_{\mathbb{S}_{\text{diff}}^{\text{ULA}}}\right]_{1} & \dots & \left[ \tilde{\boldsymbol{\mathcal{U}}}_{\mathbb{S}_{\text{diff}}^{\text{ULA}}}\right]_{2}  \\
\begin{array}{r} \vdots \end{array} &\begin{array}{r} \vdots \end{array}&  \ddots & \begin{array}{r} \vdots \end{array}\\
\left[\tilde{\boldsymbol{\mathcal{U}}}_{\mathbb{S}_{\text{diff}}^{\text{ULA}}}\right]_{2L-1} & \left[ \tilde{\boldsymbol{\mathcal{U}}}_{\mathbb{S}_{\text{diff}}^{\text{ULA}}}\right]_{2L-2} & \dots & \left[ \tilde{\boldsymbol{\mathcal{U}}}_{\mathbb{S}_{\text{diff}}^{\text{ULA}}}\right]_{L}
\end{array}\right), \nonumber
\end{eqnarray}
where $L = (|\mathbb{S}_{\text{diff}}^{\text{ULA}}| +1)/2 = MN+1$ and  
\begin{eqnarray}
[\tilde{\boldsymbol{\mathcal{U}}}_{\mathbb{S}_{\text{diff}}^{\text{ULA}}}]_l = \frac{1}{\mathbb{L}(l)} \sum_{n_1,n_2 \in \mathbb{L}(l)}^{} [\tilde{\textbf{R}}_{\boldsymbol{\mathcal{U}}}]_{n_1,n_2},
\end{eqnarray}
where $\tilde{\textbf{R}}_{\boldsymbol{\mathcal{U}}} = \frac{1}{T}\sum_{i = 1}^{T} \boldsymbol{\mathcal{U}}(t_i) \boldsymbol{\mathcal{U}}^H(t_i)$ is the sample covariance matrix and $\mathbb{L}(l)$ is defied as
\begin{eqnarray}
\mathbb{L}(l)= \{(n_1,n_2) \in \mathbb{S}^2: n_1 - n_2 = l, l \in \mathbb{S}_{\text{diff}}^{\text{ULA}}   \}.
\end{eqnarray} In other words, $\mathbb{L}(l)$ is set of  the pairs $(n_1,n_2)$ that has contribution to the contiguous part of the co-array $\mathbb{S}_{\text{diff}}^{\text{ULA}}$ with index $l$. Note that $\textbf{R}_{\boldsymbol{\mathcal{U}}-\text{SS}} \in \mathbb{C}^{(MN+1)\times (MN+1)}$ provides the same DOA estimation performance as compared to the conventional smoothed covariance matrix computed in \cite{coprimeDSPConf} for finite snapshot case  \cite{spatialSmoothingRemark}. Once $\textbf{R}_{\boldsymbol{\mathcal{U}}-\text{SS}}$ and $\textbf{R}_{\boldsymbol{\mathcal{V}}-\text{SS}}$ are computed they are inserted into the MUSIC algorithm to obtain the MUSIC pseudo-spectra as
\begin{align}
\label{musicSpectrumY}
P_{u}(\varphi) =& \frac{1}{\textbf{a}_{u}^H(\varphi) \textbf{E}_{{u}_n} \textbf{E}_{{u}_n}^H \textbf{a}_{u}(\varphi)  } ,\\
\label{musicSpectrumZ}
P_{{v}}(\vartheta) =& \frac{1}{\textbf{a}_{v}^H(\vartheta) \textbf{E}_{{v}_n} \textbf{E}_{{v}_n}^H \textbf{a}_{v}(\vartheta)  },
\end{align}
where $\varphi,\vartheta$ are associate DOA angles and they are defined as
\begin{align}
\varphi =& - \sin(\phi)\sin(\frac{\Omega}{2}) + \sin(\theta)\cos(\frac{\Omega}{2}) ,\\
\vartheta =&  \sin(\phi)\sin(\frac{\Omega}{2}) + \sin(\theta)\cos(\frac{\Omega}{2}),
\end{align}
where $\Omega$ is known before DOA estimation procedure. $\textbf{a}_{u}(\varphi)$ and $\textbf{a}_{v}(\vartheta)$ are the steering vectors constructed by using the position sets $u_i, v_i \in \mathbb{S}_{\text{diff}}^{\text{ULA-SS}}$ where
\begin{eqnarray}
\mathbb{S}_{\text{diff}}^{\text{ULA-SS}} = \{nd: 0\leq n \leq MN \}.
\end{eqnarray}
$\textbf{E}_{{u}_n}$ and $\textbf{E}_{{v}_n}$ are the noise subspace eigenvector matrices of $\textbf{R}_{\boldsymbol{\mathcal{U}}-\text{SS}}$ and $\textbf{R}_{\boldsymbol{\mathcal{V}}-\text{SS}}$ respectively.
Once $P_u(\varphi)$ and $P_v(\vartheta)$ are computed, the associate DOA angles $\{ \varphi_k,\vartheta_k \}_{k=1}^K$ are found from the highest peaks of   $P_u(\varphi)$ and $P_v(\vartheta)$. Then the estimated DOA angles are obtained as
\begin{align}
\label{theta}
\hat{\theta}_k &= {\sin}^{-1}\{\frac{\hat{\varphi}_k + \hat{\vartheta}_k}{2\cos(\Omega/2)}\}, \\
\label{phiTrans}
\hat{\phi}_k &= \sin^{-1}\{\frac{\hat{\vartheta}_k - \sin(\hat{\theta}_k)\cos(\Omega/2)}{\sin(\Omega/2)}\}.
\end{align}

\begin{figure}[t]
	\centering
	{\includegraphics[width=.37\textheight]{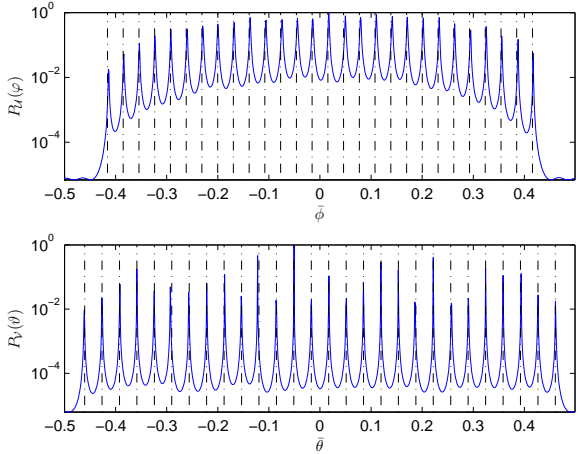} } 
	\caption{The azimuth (Top) and elevation (Bottom) estimation performance of VCA with $M = 4$, $N=7$, $M_{\text{VCA}} = 27$. The number of sensors in each portion is $2M + N-1 = 14$, $\Omega = 53.1513^{\circ}$ and number of sources is $K = 28$, SNR $ = 0$dB and the number of snapshots is $T = 1000$. The vertical lines denote the true source locations in terms of $\varphi$ and $\vartheta$. The source locations are equally spaced in the following intervals $\phi\in [-0.49,0.49]$ and $\theta\in [-0.05,0.05]$, hence $\vartheta\in [-0.4159,0.4159]$ and $\varphi\in [-0.4606,0.4606]$ The horizontal axes represent the normalized spectrum with  $\bar{\phi} = \sin(\varphi)$ and $\bar{\theta} = \sin(\vartheta)$.}
	\label{figAzElEst}
\end{figure}

In Fig. \ref{figAzElEst}, the line spectra for  $P_{u}(\varphi)$ and  $P_{{v}}(\vartheta)$ is presented by using $\textbf{R}_{\boldsymbol{\mathcal{U}}-\text{SS}}$ and $\textbf{R}_{\boldsymbol{\mathcal{V}}-\text{SS}}$ in the MUSIC algorithm. While $P_{u}(\varphi)$ and  $P_{{v}}(\vartheta)$ provide peaks at true source locations, the estimated azimuth and elevation angles are not paired due to 1-D searches. In order to obtain a paired estimation results, the cross-covariance matrix of two portions of VCA is utilized in the following section for accurate 2-D DOA estimation.
\section{2-D Paired DOA Estimation With VCA}
\label{secPaired2D}
In order to obtain the paired DOA estimates, the cross-covariance of $\boldsymbol{\mathcal{U}}(t_i)$ and $\boldsymbol{\mathcal{V}}(t_i)$ is computed. In the following, we first discuss the estimation of the azimuth angles by using only $\boldsymbol{\mathcal{U}}(t_i)$. Then the elevation angles are estimated which are automatically paired with the estimated azimuth angles. 
\subsection{Azimuth Angle Estimation}
\label{secAzEst}
The MUSIC pseudo-spectrum given in (\ref{musicSpectrumY}) and (\ref{musicSpectrumZ}) are used to for the estimation of azimuth angles. $\{\hat{\vartheta}_k,\hat{\varphi}_k\}_{k=1}^K$ can be obtained from the highest peaks of $P_{u}(\varphi)$ and$P_v(\vartheta)$. Using the transformation in (\ref{phiTrans}), azimuth angles can be estimated. In order to obtained paired elevation angles, the array steering matrix $\hat{\textbf{A}}_{u} \in \mathbb{C}^{(2M+N -1) \times K}$ can be constructed as $\hat{\textbf{A}}_{u} = [ \textbf{a}_{u}(\hat{\varphi}_1),\dots,  \textbf{a}_{u} (\hat{\varphi}_K)]$. In the next part, $\hat{\textbf{A}}_{u}$ will be used  for elevation angle estimation.
 
\subsection{Elevation Angle Estimation}
In order to estimate the elevation angles the cross-covariance matrix is computed as
\begin{eqnarray}
\textbf{R}_{\boldsymbol{\mathcal{U}}\boldsymbol{\mathcal{V}}} = E \{\boldsymbol{\mathcal{U}}(t) \boldsymbol{\mathcal{V}}^H(t)  \} = \textbf{A}_{u} \textbf{R}_{\text{S}} \textbf{A}_{v}^H,
\end{eqnarray}
where the noise terms are vanished due to the assumption that the noise is spatially white. Note that in practice, the sample cross-covariance matrix $\hat{\textbf{R}}_{{\boldsymbol{\mathcal{U}}\boldsymbol{\mathcal{V}}}}= \frac{1}{T}\sum_{i=1}^{T} \boldsymbol{\mathcal{U}}(t_i)\boldsymbol{\mathcal{V}}^H(t_i)$ is available and the noise terms are very small. Now our aim is to estimate the steering matrix $\textbf{A}_{v}$ whose columns correspond to the elevation angles which are paired with the columns of the estimated steering matrix $\hat{\textbf{A}}_{v}$. Since the columns of ${\textbf{A}}_{u}$ and $\textbf{A}_{v}$ have the same order, this process will yield an automatically paired azimuth and elevation angle estimates.

\begin{figure}[t]
	\centering
	\subfloat[]{\includegraphics[width=.37\textheight]{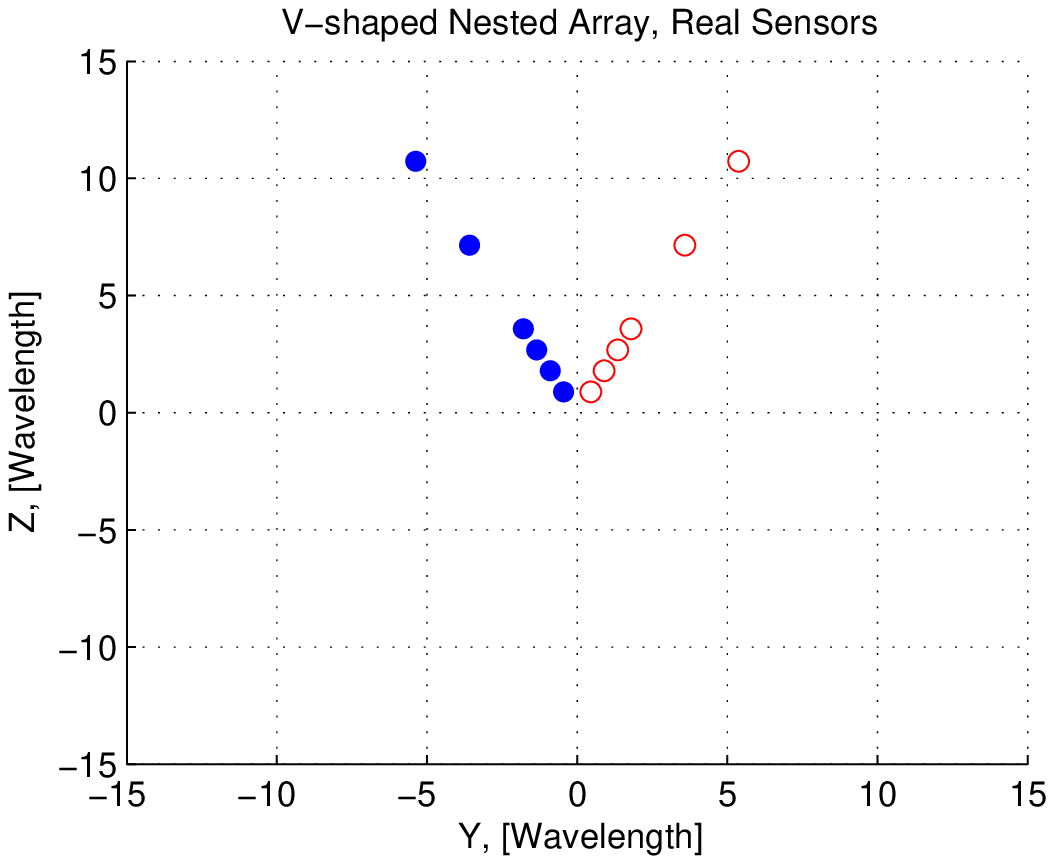}%
		\label{figPosNesteda} }  \\
	\subfloat[]{\includegraphics[width=.37\textheight]{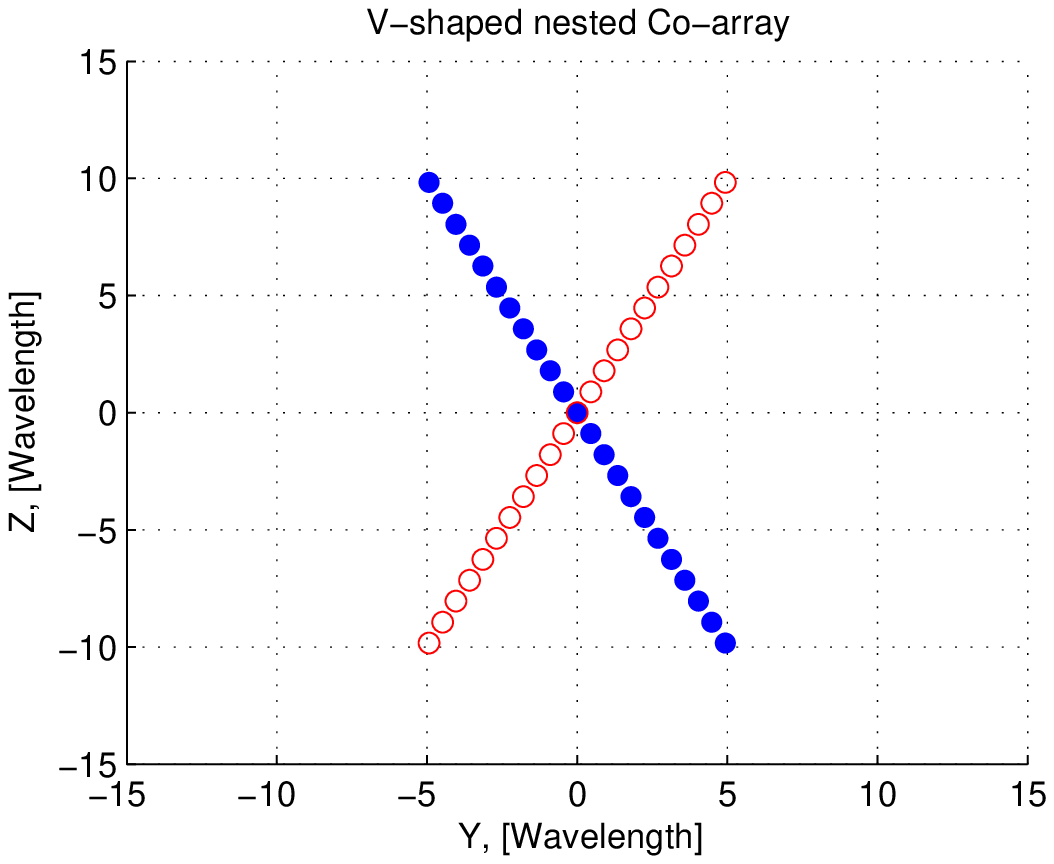}%
		\label{figPosNestedb}	}
	\caption{V-shaped nested array (VNA) structure for $N_1 = N_2 = 3$. (a) The real sensor positions. (b) Co-array of each portion of VNA.}
	\label{figPosNested}
\end{figure} 
Hence we solve the following least squares problem, i.e.
\begin{eqnarray}
\label{AzLS}
\hat{\textbf{A}}_{v} = \arg \min_{\textbf{A}_{v}} || \textbf{R}_{{\boldsymbol{\mathcal{U}}\boldsymbol{\mathcal{V}}}} - \hat{\textbf{A}}_{u} {\textbf{R}}_{\text{S}} \textbf{A}_{v}^H ||_F^2,
\end{eqnarray}
where the knowledge of ${\textbf{R}}_{\text{S}}$ is required for the computation of $\textbf{A}_{v}$. In order to estimate $\textbf{R}_{\text{S}}$ we consider the eigendecomposition of the  covariance matrix $\textbf{R}_{{\boldsymbol{\mathcal{U}}}}$ in (\ref{Ry}) as
\begin{eqnarray}
\label{Ry2}
{\textbf{R}}_{{\boldsymbol{\mathcal{U}}}}  = \textbf{E}_{u} \boldsymbol{\Lambda} \textbf{E}_{u}^H,
\end{eqnarray}
where $\textbf{E}_{u} = [\textbf{E}_{{u}_s}\text{ } \textbf{E}_{{u}_n} ]$ and $\textbf{E}_{{u}_s}$, $\textbf{E}_{{u}_n}$ are the signal and noise subspace eigenvector matrices respectively. $\boldsymbol{\Lambda}$ is a diagonal matrix composed of the eigenvalues of $\textbf{R}_{{\boldsymbol{\mathcal{U}}}}$. (\ref{Ry2}) can also be written as
\begin{eqnarray}
\label{Ry3}
\textbf{R}_{{\boldsymbol{\mathcal{U}}}} = \textbf{E}_{{u}_s}\boldsymbol{\Lambda}_s \textbf{E}_{{u}_s}^H + \textbf{E}_{{u}_n} \boldsymbol{\Lambda}_n \textbf{E}_{{u}_n}^H,
\end{eqnarray}
where $\boldsymbol{\Lambda}_s\in \mathbb{C}^{K\times K}$ and $\boldsymbol{\Lambda}_n\in \mathbb{C}^{(2M+N-1-K)\times( 2M + N- 1-K)}$ are diagonal matrices composed of the eigenvalues of $\textbf{R}_{{\boldsymbol{\mathcal{U}}}}$ with respect to signal and noise subspaces respectively. Using (\ref{Ry}), (\ref{Ry3}) and the fact that the columns of ${\textbf{A}}_{{u}}$ and $\textbf{E}_{u_s}$ span the same space, $\textbf{R}_{\text{S}}$ can be estimated  from
\begin{eqnarray}
\label{RsEst}
\hat{\textbf{R}}_{\text{S}} = \hat{\textbf{A}}_{{u}}^{\dagger} \textbf{E}_{{u}_s} \boldsymbol{\Lambda}_s \textbf{E}_{{u}_s}^H \left( \hat{\textbf{A}}_{{u}}^H\right) ^{\dagger},
\end{eqnarray}
where $ (\cdot)^{\dagger}$ denotes the Moore-Penrose pseudo-inverse operation. Then the steering matrix $\textbf{A}_{v}$ is estimated from (\ref{AzLS}) by using the following closed form expression, i.e.
\begin{eqnarray}
\label{AzEst}
\hat{\textbf{A}}_{v} = \left( \hat{\textbf{R}}_{\text{S}}^{-1}\left( \hat{\textbf{A}}_{{u}}\right) ^{\dagger}\textbf{R}_{{\boldsymbol{\mathcal{U}}\boldsymbol{\mathcal{V}}}}  \right) ^H.
\end{eqnarray}
Using (\ref{RsEst}), (\ref{AzEst}) can be written explicitly as
\begin{eqnarray}
\hat{\textbf{A}}_{v} = \left( \left( \hat{\textbf{A}}_{{u}}^{\dagger}  \textbf{E}_{{u}_s} \boldsymbol{\Lambda}_s \textbf{E}_{{u}_s}^H (\hat{\textbf{A}}_u^H)^{\dagger}\right) ^{-1} \hat{\textbf{A}}_{{u}} ^{\dagger}\textbf{R}_{{\boldsymbol{\mathcal{U}}\boldsymbol{\mathcal{V}}}}  \right)^H.
\end{eqnarray}
Note that the size of the estimated steering matrix $\hat{\textbf{A}}_{v}$ is ${(2M+N-1)\times K}$ and in underdetermined case we have $(2M+N-1) < K$. While in this case the covariance matrix of $\hat{\textbf{A}}_{v}$ does not lead to accurate results due to rank-deficiency, we instead use the columns of $\hat{\textbf{A}}_{v}$ to estimate the elevation angles one by one so that each elevation angle is paired with the corresponding azimuth angle. Hence the elevation angles can be estimated by the MUSIC algorithm using the covariance matrix $\hat{\textbf{R}}_{{\boldsymbol{\mathcal{V}}}_k} = [\hat{\textbf{A}}_{v}]_{:,k} [\hat{\textbf{A}}_{v}]_{:,k}^H$. In other words, $\hat{\textbf{R}}_{{\boldsymbol{\mathcal{V}}}_k}$ can be obtained for the $k$th column of $\hat{\textbf{A}}_{v}$. Since $\text{rank}\{ \hat{\textbf{R}}_{{\boldsymbol{\mathcal{V}}}_k}\} =1 $, 1-D MUSIC algorithm is used to estimate $\theta_k$. In particular, $\vartheta_k$ is estimated from 
\begin{eqnarray}
\label{estTheta}
\hat{\vartheta}_k = \arg \max_{\vartheta} \frac{1}{\textbf{a}_{v}^H(\vartheta)\textbf{G}_k\textbf{G}_k^H \textbf{a}_{v}(\vartheta)},
\end{eqnarray}
for $k = 1\dots,K$ where $\textbf{a}_{v}(\vartheta) \in \mathbb{C}^{(2M+N -1)}$ is the steering vector corresponding to the position set $v_i \in \mathbb{S}$. $\textbf{G}_k\in \mathbb{C}^{(2M+N -1) \times (2M+N -2)}$ is the noise subspace eigenvector matrix of $\hat{\textbf{R}}_{{\boldsymbol{\mathcal{V}}}_k}$. Once $\hat{\vartheta}_k$ is obtained, the elevation angle $\hat{\theta}_k$ can be found from (\ref{theta}).

\textit{Remark:} The proposed 2-D DOA estimation approach is based on 1-D searches in both azimuth and elevation dimensions then a pairing stage is employed. While the proposed approach can handle detecting the targets and estimate the azimuth and elevation angles correctly, it cannot distinguish the two sources whose azimuth and elevation angles are interchanged.

\section{Extension For V-Shaped Nested Arrays}
\label{secNested}
In this section, the proposed approach is extended to nested arrays. A 2-level nested array composed of two subarrays with $N_1 $ and $N_2$ elements can resolve up to $K \leq N^2$ sources where $N = N_1 + N_2$ total number of sensors in the array \cite{nestedArray}. We assume that $N_1=N_2=N/2$ for simplicity. Now we consider two nested arrays in ${\mathcal{U}}$- and ${\mathcal{V}}$-axes for 2-D DOA estimation and the total number of sensors in VNA is $M_{\text{VNA}} = 2N$. In Fig. \ref{figPosNesteda}, the positions of the sensors for a VNA is presented for $N_1 =N_2=3$. The coarray for VNA is shown in Fig. \ref{figPosNestedb} which is composed of $N$-elements in each portion and $M_{\text{VNA}} = 2N$. The number of sources that can be estimated from each portion is $ (N^2/4 + N/2 -1 )$. Again this number is much less than the total number of sensors in CPA and GCPA. In order to obtain $\Omega$, similar computation can be done and $\bar{M} = 2N +1$ for VNA. For 2-D DOA estimation the same procedure from (\ref{Ry}) to (\ref{estTheta}) can be followed as for VCA and 2-D paired DOA angles can be estimated.

\begin{table}[h]
	\caption{Computational complexity comparison.}
	\label{tableComplexity}
	\centering
	\begin{tabular}{|c||c|}
		\hline
		\hline
		VCA &  \footnotesize $ \left( MN+1\right) \left( N_{\varphi}\left( MN+1-K \right)  + KN_{\vartheta}MN\right) $ \\
		\hline
		VNA &  \footnotesize$ \left( \frac{N^2}{4} + \frac{N}{2}\right) \left( N_{\varphi}\left( \frac{N^2}{4} + \frac{N}{2}-K \right)  + KN_{\vartheta}(\frac{N^2}{4} + \frac{N}{2}-1)\right) $ \\
		\hline
		CPA &\footnotesize $ N_{\phi}N_{\theta}\frac{M_1^4}{M_2^2} +  N_{\phi}N_{\theta}\frac{M_2^4}{M_1^2}  $\\
		\hline
		GCPA &\footnotesize $ N_{\phi}N_{\theta}\left(M_1N_1(M_1N_1-K) + M_2N_2(M_2N_2-K) \right)  $\\
		\hline
	\end{tabular}
\end{table}

\section{Computational Complexity}
\label{secCompComp}
In this section, the complexity of the proposed method is discussed. In order to estimate the azimuth angles 1-D spectral search is required together with the computation of the singular value decomposition (SVD) of $\textbf{R}_{\boldsymbol{\mathcal{U}}\text{-SS}}$ to obtain the noise subspace. Hence $O\left( \left( MN+1\right) ^3 + N_{\varphi}\left( MN+1\right) \left( MN+1-K\right) \right) $ is the complexity of azimuth angle estimation where $(MN+1)^3$ is the complexity of SVD and $N_{\varphi}$ is the number of search angles in the grid. In order to estimate the elevation angles the SVD of $\hat{\textbf{R}}_{\boldsymbol{\mathcal{V}}_k}$ is computed for $k=1,\dots,K$. Hence the complexity order of elevation angle estimation is $O\left( K\left( MN+1\right) ^3 + KN_{\vartheta}\left( MN+1\right) MN\right) $ where $N_{\vartheta}$ is the number of search angles in the elevation grid. Due to the fact that the spectral search is much heavier burden than the other operations it suffices to state the complexity of the proposed method as  $O \left( \left( MN+1\right) \left( N_{\varphi}\left( MN+1-K \right)  + KN_{\vartheta}MN\right) \right) $ where we ignore the other terms. In order compare the complexity of the proposed method with GCPA and CPA we note the following. GCPA and CPA  uses $M_1\times N_1$, $M_2\times N_2$ and $M_1\times M_1$, $M_2\times M_2$ arrays respectively which require much higher number of sensors than VCA or VNA. Another disadvantage of GCPA and CPA is to use 2-D search algorithms which require $N_{\phi}N_{\theta}$ grid points to compute the MUSIC pseudo-spectrum. The complexity of the algorithm in \cite{LshapedCoprimeGeneralizedCL2017} using GCPA is $O\left( N_{\phi}N_{\theta}\left(M_1N_1(M_1N_1-K) + M_2N_2(M_2N_2-K) \right) \right) $ which is much higher than the complexity of the proposed method due to 2-D search. The complexity of the method in \cite{Coprime2DPlanarSJ} for CPA is $O\left( N_{\phi}N_{\theta}\frac{M_1^4}{M_2^2} +  N_{\phi}N_{\theta}\frac{M_2^4}{M_1^2} \right) $ which is also much higher than the complexity of the proposed method. In Table \ref{tableComplexity}, the complexities of the arrays are summarized.

\section{Numerical Simulations}
\label{Simulations}
In this section, the performance of the proposed method is evaluated with numerical simulations. We show the maximum number of resolvable sources of the arrays VCA, VNA, CPA and GCPA in Table \ref{tableResolvability}. As it is seen VCA and VNA require much less number of sensors to resolve the same number of sources. The improvement is attributed to the use of sparse sampling and the larger aperture of virtual arrays. In Table \ref{tableNumberOfSensors}, the number of sensors of each array is presented. We can observe from Table \ref{tableNumberOfSensors} that the proposed array structures have less number of elements as compared to the other array geometries such as CPA \cite{Coprime2DPlanarSJ} and GCPA \cite{LshapedCoprimeGeneralizedCL2017}. In the light of this information, different scenarios are considered for different number of sources. The results for required resolvability and total number of sensors are given in Table \ref{tableComparison} where the total number of the sensors in the arrays, VCA, VNA, CPA and GCPA are denoted as $M_{\text{VCA}}$, $M_{\text{VNA}}$, $M_{\text{CPA}}$ and $M_{\text{GCPA}}$ respectively. As it is seen, the proposed array geometries, VCA and VNA, provide less number of sensors in all scenarios considered. Moreover, VNA provides the least number of sensors to resolve source directions in underdetermined case. The other planar array geometries CPA and GCPA require much more sensors since they do not exploit the sparsity property as VCA and VNA.

\begin{table}[h]
	\caption{Maximum number of sources resolvable for different arrays.}
	\label{tableResolvability}
	\centering
	\begin{tabular}{|c||c|}
		\hline
		\hline
		VCA &  $ MN$ \\
		\hline
		VNA &  $N^2/4 + N/2-1$ \\
		\hline
		CPA & $ \text{min}\{M_1^2,M_2^2\}$\\
		\hline
		GCPA & $\text{min}\{M_1N_1,M_2N_2\}$ \\
		\hline
	\end{tabular}
\end{table}

\begin{table}[h]
	\caption{Number of sensors for different arrays.}
	\label{tableNumberOfSensors}
	\centering
	\begin{tabular}{|c||c|}
		\hline
		\hline
		VCA &  $4M + 2N -3 $  \\
		\hline
		VNA &  $2N$\\
		\hline
		CPA & $ M_1^2+M_2^2$\\
		\hline
		GCPA & $M_1N_1+M_2N_2$  \\
		\hline
	\end{tabular}
\end{table}

\begin{figure*}[t!]
	\centering
	\subfloat[]{\includegraphics[width=.24\textheight]{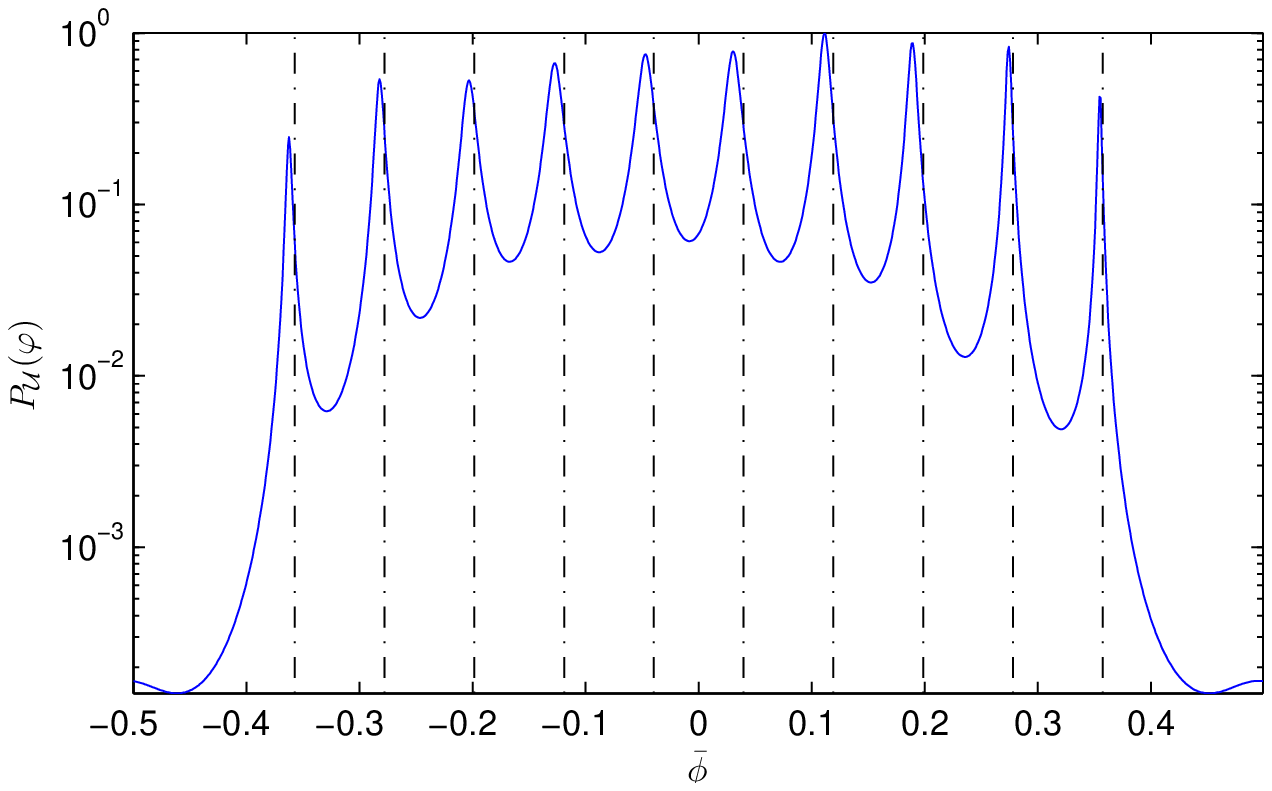}%
		\label{figAzSpec} }
	\subfloat[]{\includegraphics[width=.24\textheight]{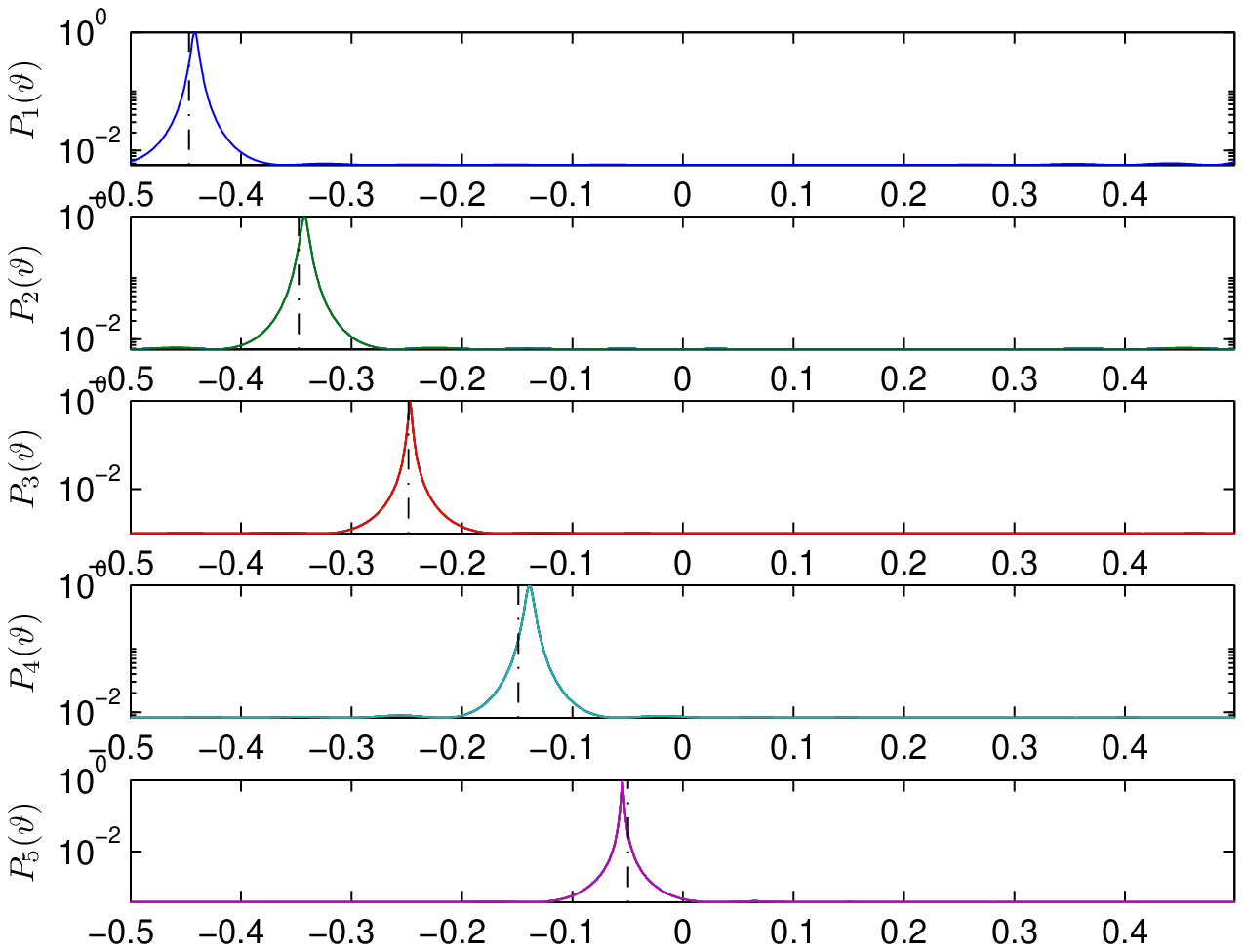}%
		\label{figElSpec1} } 
	\subfloat[]{\includegraphics[width=.24\textheight]{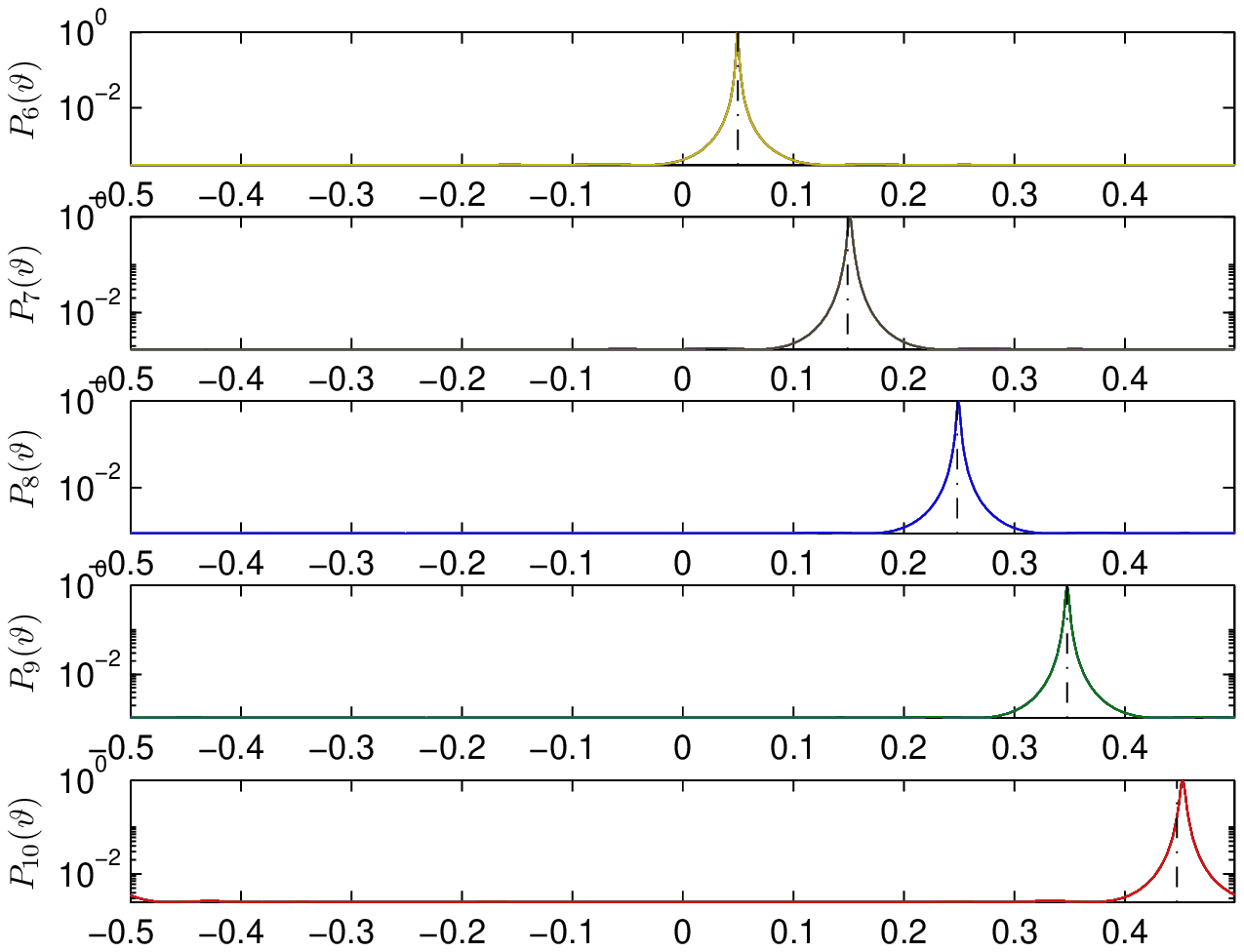}%
		\label{figElSpec2}	}
	\caption{Azimuth (a) and elevation (b)-(c) spectrums for VCA with $M = 2$, $N = 5$, $M_{\text{VCA}} = 15$. The number of sensors in each portion is $2M + N -1 = 8$. The number of sources is $K = 10$. $\Omega= 53.2856^{\circ}$, SNR =$0$ dB and $T = 1000$. The vertical lines denote the true source locations in terms of $\varphi$ and $\vartheta$. The source locations are equally spaced in the following intervals $\phi\in [-0.45,0.45]$ and $\theta\in [-0.1,0.1]$, hence $\vartheta\in [-0.4471,0.4471]$ and $\varphi\in [-0.3574,0.3574]$ The horizontal axes represent the normalized spectrum with  $\bar{\phi} = \sin(\varphi)$ and $\bar{\theta} = \sin(\vartheta)$.}
	\label{figAzElSpec}
\end{figure*}

In Fig. \ref{figAzElSpec}, the line spectrums for azimuth (a) and elevation (b)-(c) are presented where $M = 2$, $N = 5$, $M_{\text{VCA}} = 15$. The number of sensors in each portion is $2M + N -1 = 8$. The number of sources is $K = 10$. $\Omega= 53.2856^{\circ}$, SNR = $0$dB and $T = 1000$. As it is seen, the proposed method can handle resolving $MN=10$ sources with using $8$ sensor in each portion and total $M_{\text{VCA}} =15$ sensor elements.

\begin{table*}[!h]
	\caption{Comparison of array structures for different scenarios.}
	\label{tableComparison}
	\centering
	\begin{tabular}{|c||c|c|c|c|}
		\hline
		\hline
		$K$	&VCA &  VNA & CPA & GCPA \\
		\hline
		8& \footnotesize$ M=2$, $N = 5$, $M_{\text{VCA}} = 15$ &\footnotesize  $N = 6$, $M_{\text{VNA}} = 12$ &\footnotesize $M_1 = 3$, $M_2=4$, $M_{\text{CPA}} = 12$&\footnotesize $M_1= 3$, $N_1=3$, $M_2=2$, $N_2=4$,  $M_{\text{GCPA}} = 15$ \\
		\hline
		10& \footnotesize$ M=2$, $N = 5$, $M_{\text{VCA}} = 15$ &\footnotesize  $N = 6$, $M_{\text{VNA}} = 12$ &\footnotesize $M_1 = 4$, $M_2=5$, $M_{\text{CPA}} = 40$&\footnotesize $M_1= 4$, $N_1=4$, $M_2=3$, $N_2=5$,  $M_{\text{GCPA}} = 31$ \\
		\hline
		12& \footnotesize$ M=3$, $N = 5$, $M_{\text{VCA}} = 19$ & \footnotesize $N = 8$, $M_{\text{VNA}} = 16$ &\footnotesize $M_1 = 4$, $M_2=5$, $M_{\text{CPA}} = 40$&\footnotesize $M_1= 4$, $N_1=4$, $M_2=3$, $N_2=5$,  $M_{\text{GCPA}} = 31$ \\
		\hline
		17&\footnotesize $ M=2$, $N = 9$, $M_{\text{VCA}} = 23$ &\footnotesize  $N = 8$, $M_{\text{VNA}} = 16$ &\footnotesize $M_1 = 5$, $M_2=6$, $M_{\text{CPA}} = 60$&\footnotesize $M_1= 4$, $N_1=5$, $M_2=3$, $N_2=7$,  $M_{\text{GCPA}} = 40$ \\
		\hline
		20&\footnotesize $ M=4$, $N = 5$, $M_{\text{VCA}} = 23$ & \footnotesize $N = 10$, $M_{\text{VNA}} = 20$ & \footnotesize$M_1 = 5$, $M_2=6$, $M_{\text{CPA}} = 60$&\footnotesize $M_1= 4$, $N_1=5$, $M_2=3$, $N_2=7$,  $M_{\text{GCPA}} = 40$ \\
		\hline
		28&\footnotesize $ M=4$, $N = 7$, $M_{\text{VCA}} = 27$ &\footnotesize  $N = 10$, $M_{\text{VNA}} = 20$ & \footnotesize$M_1 = 6$, $M_2=7$, $M_{\text{CPA}} = 84$& \footnotesize$M_1= 4$, $N_1=7$, $M_2=5$, $N_2=6$,  $M_{\text{GCPA}} =57$ \\
		\hline
	\end{tabular}
\end{table*}

\begin{figure}[h]
	\centering
	\includegraphics[width=.35\textheight]{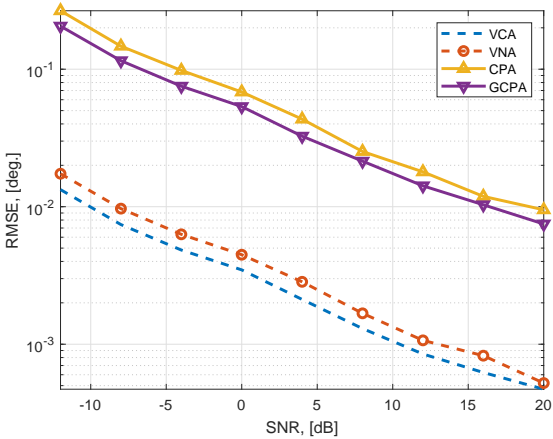}
	\caption{Performance comparison of different array geometries. The number of sensors in the arrays are  $M_{\text{VCA}} = 39$, $M_{\text{VNA}} = 39$, $M_{\text{CPA}} = 40$ and $M_{\text{GCPA}} = 39$ respectively. There are $K=6$ sources which are equally spaced in the following intervals $\phi\in [-0.45,0.45]$ and $\theta\in [-0.1,0.1]$ and $T = 500$. }
	\label{figSNRTest}
\end{figure}

\begin{figure}[h]
	\centering
	\includegraphics[width=.35\textheight]{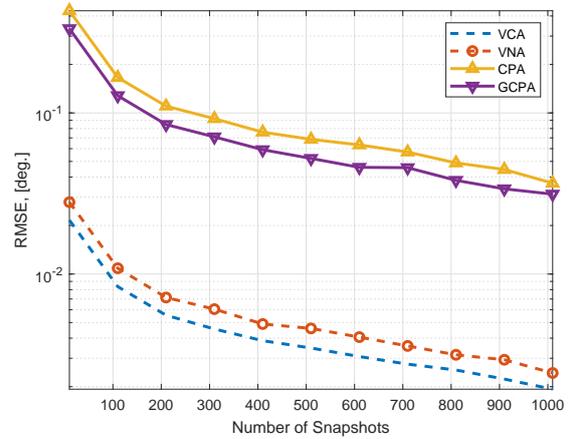}
	\caption{DOA estimation performance vs number of snapshots when SNR=0dB. The number of sensors in the arrays are  $M_{\text{VCA}} = 39$, $M_{\text{VNA}} = 39$, $M_{\text{CPA}} = 40$ and $M_{\text{GCPA}} = 39$ respectively. There are $K=6$ sources which are equally spaced in the following intervals $\phi\in [-0.45,0.45]$ and $\theta\in [-0.1,0.1]$ and $T = 500$. }
	\label{figSnapTest}
\end{figure}

In Fig. \ref{figSNRTest}, the estimation performance of VCA, VNA, CPA and GCPA are compared. Note that the RMSE calculation is performed in terms of both azimuth and elevation as 
\begin{eqnarray}
\text{RMSE}_{\text{DOA}} = \left( {\frac{1}{2JK}\sum_{j = 1}^{J} \sum_{k = 1}^{K }  \left( \phi_{k} - \hat{\phi}_{k,j} \right)^2 + \left( \theta_{k} - \hat{\theta}_{k,j} \right)^2   }\right) ^{1/2} \nonumber
\end{eqnarray}
where $J$ is the number of Monte Carlo experiments and $J=100$ is selected. For a fair comparison, the number of sensors of the arrays are selected closely as $M_{\text{VCA}} = 39$, $M_{\text{VNA}} = 39$, $M_{\text{CPA}} = 40$ and $M_{\text{GCPA}} = 39$. Note that this selection is sufficient to demonstrate the performance of the considered array geometries. While CPA has one more sensor than GCPA, it still performs poorer due to its lack of array aperture. In this scenario there are $K=6$ sources and $T = 500$. As seen from Fig. \ref{figSNRTest}, VCA and VNA have superior performance as compared to the coprime planar arrays CPA and GCPA. VNA has better precision as compared to VCA since VNA has larger DOF and hence larger virtual aperture. In Fig.~\ref{figSnapTest}, DOA estimation performance comparison is presented for different number of snapshots when SNR=0dB. We obtain similar observations as seen in Fig.~\ref{figSnapTest} where VCA and VNA have superior DOA estimation performance as compared to CPA and CGPA structures.

\begin{figure}[!h]
	\centering
	\subfloat[]{\includegraphics[width=.35\textheight,height=.17\textheight]{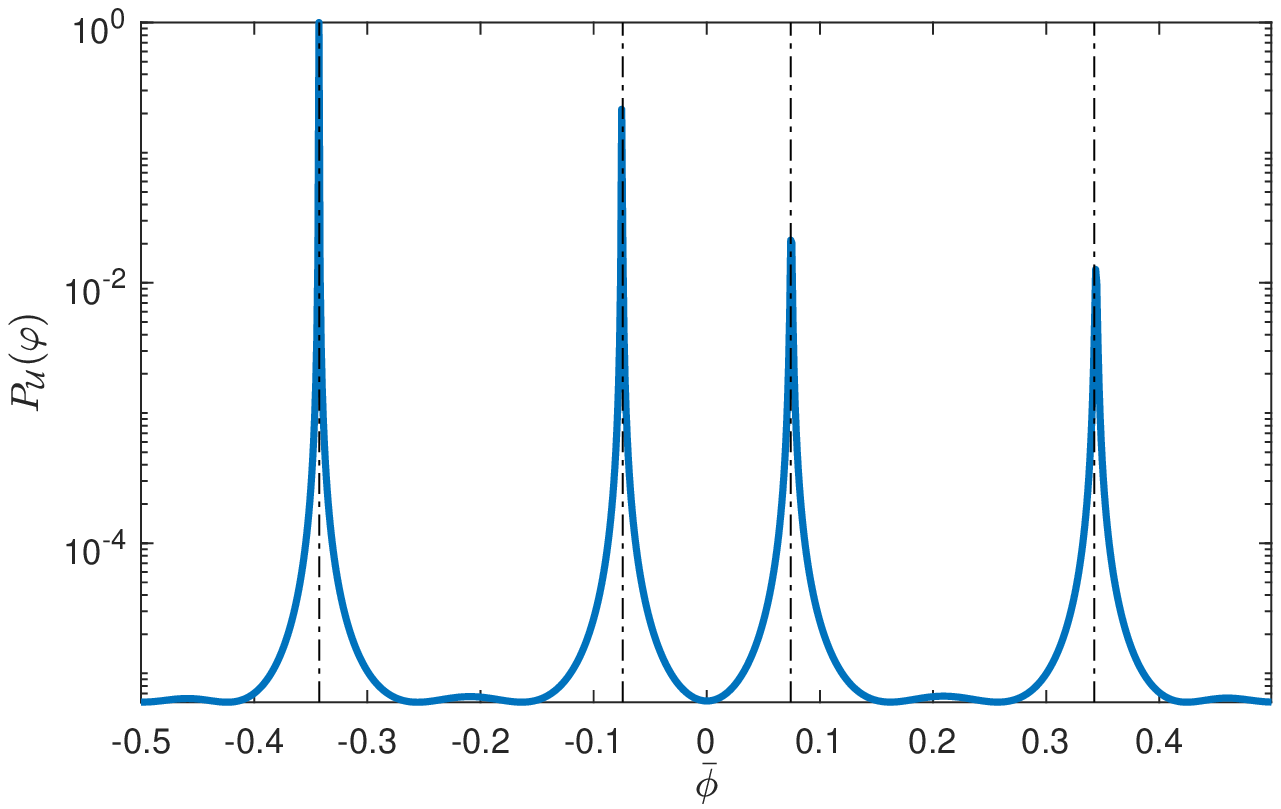}%
		\label{figAzSpecNested} } \\
	\subfloat[]{\includegraphics[width=.35\textheight]{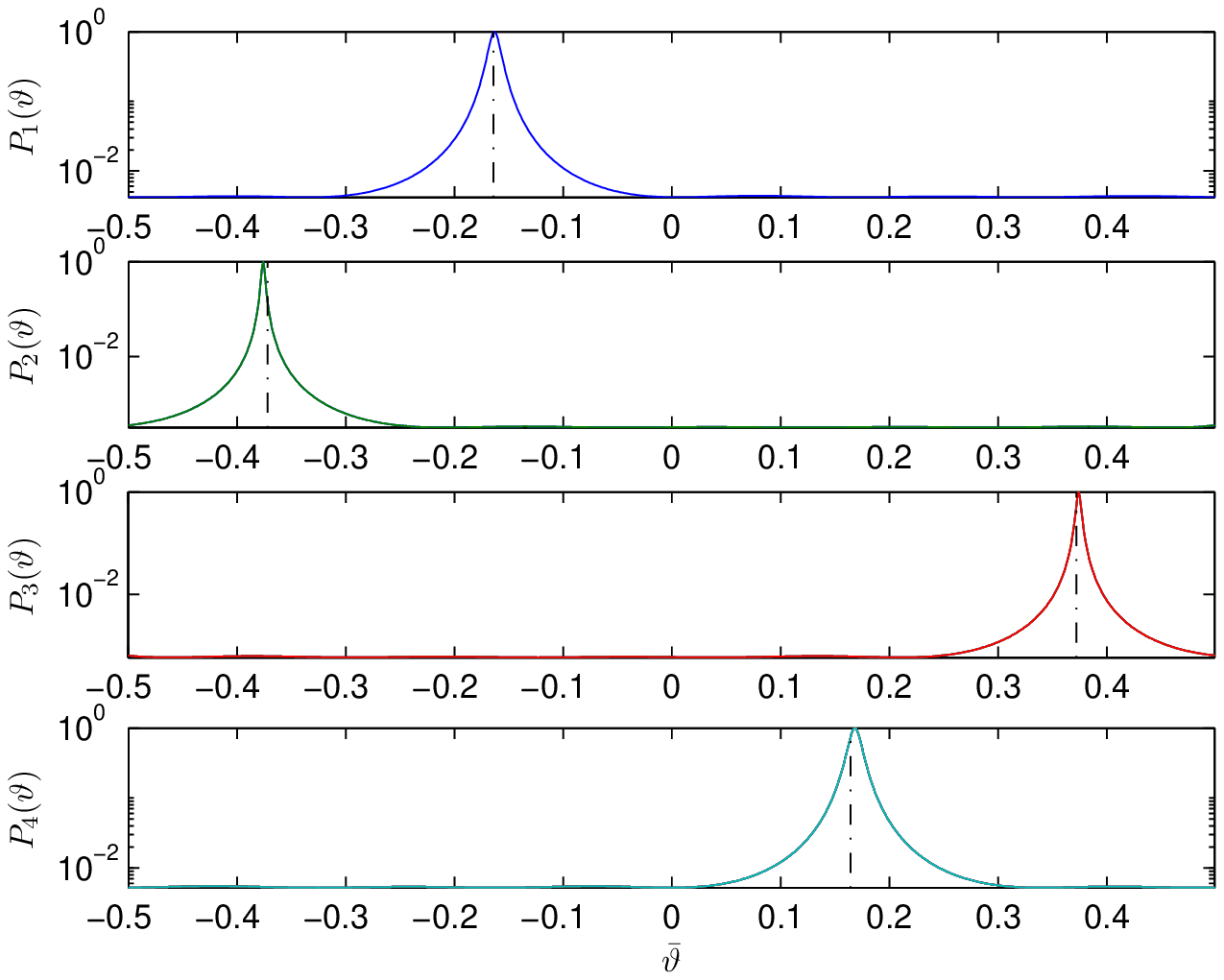}%
		\label{figElSpec1Nested} }
	\caption{Azimuth (a) and elevation (b) spectrums for VNA with $N = 4$, $M_{\text{VNA}} = 8$. The number of sensors in each portion is $N = 4$. The number of sources is $K = 4$. SNR =$0$dB and $T = 1000$. The vertical lines denote the true source locations in terms of $\varphi = [-0.0738, -0.3418, 0.3418, 0.0738]$ and $\vartheta=[-0.1641, -0.3719,	0.3719, 0.1641]$.}
	\label{figAzElSpecNested}
\end{figure}

In Fig. \ref{figAzElSpecNested}, azimuth (a) and elevation (b) spectra is given for VNA. There are $K$ sources with amplitudes $1, 0.7,0.4$ and  $0.3$ respectively and $N=4$ sensors in each portion and $M_{\text{VNA}}= 8$. The azimuth sources are located as $[-0.1000, -0.0333, 0.0333, 0.1000]$ and the elevation angles are selected as $[-0.1333, -0.4000, 0.4000, 0.1333]$. Note that the source angles are not selected in increasing fashion. Hence this experiment evaluates the pairing performance of the proposed method. As it is seen from the figure, the proposed method effectively estimates the source DOAs and accurately pairs them. 

\section{Conclusions}
\label{secConc}
In this paper, a new array geometry, namely V-shaped sparse array (VSA), is proposed as a promising structure for 2-D DOA estimation. The proposed array geometry is very efficient in terms of required real sensor elements for a certain scenario and it does not require 2-D grid search. Hence it has a low computational complexity. The sparsity of the array is exploited for both coprime and nested spatial sampling cases and their resolvability performances are evaluated. The proposed array structures VCA and VNA can resolve both azimuth and elevation angles up to $K\leq MN$ and $K \leq N^2/4 + N+2 -1$ sources respectively. These resolvability limits are much higher than the conventional planar coprime array structures such as CPA and GCPA which are recently proposed in the literature. In case of interchanged azimuth and elevation angles of two sources, the proposed method is unable to pair the sources while it can estimate the angles accurately. In future works, we concentrate on overcoming the pairing problem for this scenario.

\bibliographystyle{plain}
\bibliography{references_028}

\end{document}